# Collective phenomena in crowds - where pedestrian dynamics need social psychology

Anna Sieben[1], Jette Schumann[2], Armin Seyfried[2,3,*]

[1] Chair of Social Theory and Social Psychology, Ruhr-Universität Bochum, Universitätsstraße 150, 44801 Bochum, Germany

[2] Jülich Supercomputing Centre, Forschungszentrum Jülich, Wilhelm-Johnen-Straße, 52428 Jülich, Germany

[3] School of Architecture and Civil Engineering, University of Wuppertal, Pauluskirchstraße 7, 42285 Wuppertal, Germany

# Abstract

This article is on collective phenomena in pedestrian dynamics during the assembling and dispersal phases of gatherings. To date pedestrian dynamics have been primarily studied in the natural and engineering sciences. Pedestrians are analyzed and modeled as driven particles revealing self-organizing phenomena and complex transport characteristics. However, pedestrians in crowds also behave as living beings according to stimulus-response mechanisms or act as human subjects on the basis of social norms, social identities or strategies. To show where pedestrian dynamics need social psychology in addition to the natural sciences we propose the application of three categories – phenomena, behavior and action. They permit a clear discrimination between situations in which minimal models from the natural sciences are appropriate and those in which sociological and psychological concepts are needed.

To demonstrate the necessity of this framework, an experiment in which a large group of people (n=270) enters a concert hall through two different spatial barrier structures is analyzed. These two structures correspond to everyday situations such as boarding trains and access to immigration desks. Methods from the natural and social sciences are applied. Firstly, physical measurements show the influence of the spatial structure on the dynamics of the entrance procedure. Density, waiting time and speed of progress show large variations. Secondly, a questionnaire study (n=60) reveals how people perceive and evaluate these entrance situations. Markedly different expectations, social norms and strategies are associated with the two spatial structures. The results from the questionnaire study do not always conform to objective physical measures, indicating the limitations of models which are based on objective physical measures alone and which neglect subjective perspectives.

# Introduction

Numerous academic disciplines focus on situations in which large numbers of people use a common space and timeframe – known as gatherings. People at gatherings appear in physics as self-driven particles, in traffic engineering as pedestrians, in computer science as autonomous agents, in social psychology as groups and in sociology as collective actors – to name but a few examples. Therefore notions such as crowd, collective behavior, mass panic or mob convey diverse meanings. The need for interdisciplinary work is indicated by the many papers in the natural sciences that mention social or psychological aspects [1,2,3,4,5]. However, integrative and interdisciplinary frameworks have only rarely been developed [6,7]. This article aims at combining a natural scientific perspective on the dynamics of pedestrians with sociological and psychological concepts of action and behavior. In the following, these fields are roughly outlined.

Crowd psychology first comes to mind when searching for a psychological perspective on gatherings. However, those who made the term crowd psychology well known at the end of the 19[th] century and beginning of the 20[th] century – most prominently Gustave Le Bon – make crowd psychology unpopular nowadays. Their concept of the collective mind and their negative perspective of the crowd (as being primarily emotional, irrational, anti-social and politically dangerous) have been shown to be empirically inappropriate and conceptually problematic [8,9]. By and large, crowds do not behave in an irrational and anti-social way. On the contrary, large groups of people usually move in a quite orderly and cooperative manner, assistance is frequently given to others, even in dangerous situations and among strangers [10,11,12,13].

In contemporary social psychology, gatherings are often scaled down to phenomena observed in (small) groups. Recently, group phenomena have been increasingly studied in relation to motions and dynamics of pedestrians [14,15,16]. Large gatherings usually consist of small groups or couples and not of single individuals. Membership of a group is a powerful predictor of the tendency to help and to "stick together" [17]. Drury and colleagues argue that crowd models need to pay attention to group identities [14,65]. Furthermore, cognitive psychologists are particularly interested in wayfinding and the perception of direction signs and announcements [7,18].

Sociology has addressed collective or crowd behavior in complex social settings such as political movements. Sociologists tend to look at motivational and situational factors to explain why people start collective behavior. Furthermore, McPhail, Schweingruber and their colleagues developed an in-depth observation and analysis scheme for collective action at gatherings [11,12,13]. Sociological analysis does not focus on pedestrian dynamics – the actual movement of participants.

Physicists describe pedestrians as self-driven particles with complex interactions focusing on collective phenomena emerging as spatio-temporal structures in many-particle systems. For an overview, we refer to the proceedings of the conferences series 'Traffic and Granular Flow' [19,20] and the review article [21,22]. Traffic and safety engineers deal with pedestrians as a traffic mode and use concepts like density and flow to describe the transport properties of pedestrian facilities [23,24,25]. Typical research applications are evacuation of assembly buildings or cruise ships, design of transport infrastructure or the organization of large-scale events [26,27,28]. Further fields are computer science where agent-based models are used to model autonomous robots and agents for computer games or animated movies [29] or geography [22]. Following the principle of Occam's razor, natural scientists seek the simplest model offering a sufficiently precise description of the system. One class of models is individual-based models describing the movement of single pedestrians interacting with other pedestrians and the environment, e.g. cellular automata [30,31,32], force-based models [1,2,3,33] or velocity models [34,35,36,37,38].

To sum up, social scientists treat pedestrians as psychological and social beings whereas physicists and engineers treat them as moving particles – resulting, in extreme cases, in mutually exclusive positions of psychological realism and physical reductionism. This paper aims at developing an integrative perspective by introducing distinctions between three classes of collectivity at gatherings: Collective phenomena, collective behavior and collective action. The following article is divided into four parts. First, the theoretical framework is developed on the basis of sociological action theory with three prominent examples, namely lane formation in pedestrian streams, flocks of animals and clogging of escaping people at bottlenecks. Second, an experiment with two different entrance setups is analyzed and the limits of models from natural science are discussed. Third, subjective strategies and social norms for both entrance setups are described on the basis of a questionnaire study. We use the empirical results to introduce the concept of collective action for entrance setups. Fourth, in the concluding section consequences for the modeling of pedestrian dynamics are discussed. We argue that the subjective perspectives of pedestrians need to be taken into account in a subset of gatherings in order to model their dynamics appropriately.

# Theoretical framework

Collectivity in gatherings can best be identified from an observer-based perspective. It involves at least two people and is identifiable by a synchronized character. Collectivity shows patterns (of movement) which go beyond the simple aggregation of individuals. It can be accomplished by different means and mechanisms. This article develops a systematic distinction between three classes of collectivity: collective phenomena, collective behavior and collective action. It draws upon the basic distinction between behavior and action as developed in sociological action theory [39,40]. It should be noted that collective action and behavior cannot be conceptualized as the opposite of individual action and behavior. Rather, collective behavior is always based upon and consists of individual behavior. But not vice versa. There are phases at gatherings during which participants simply act on their own. These phases might be called individualized. At gatherings, collective and individualized actions take turns [12].

The first class – collective phenomena – can be described with reference to physical models only. No psychological or sociological concepts are needed for their description. Collective phenomena emerge as spatio-temporal structures in cumulated trajectories of entities indicating that the movement displays more than individual characteristics. Lane formation in pedestrian streams can be classified under this category: In bidirectional streams, the formation of lanes and bands is observable leading to clusters of pedestrians moving in one direction. Figure 1 shows a snapshot of an experiment under laboratory conditions, see movie S1 and [41] for more details. Movement in opposite directions can lead to collisions, which pedestrians prevent by changing the direction and speed of their motion. With increasing density, the number of options to prevent conflicts decreases. Sorting and separation into lanes and bands leads to fewer potential conflicts and thus to fewer forced changes of speed and direction of motion. This sorting occurs unconsciously without any explicit arrangement or verbal communication.

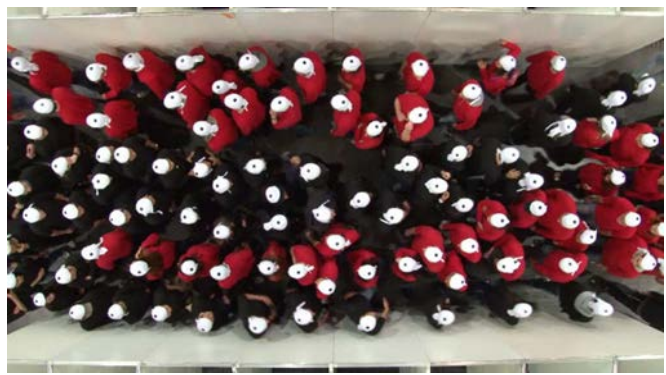

**Fig 1.** Snapshot of an experiment with bidirectional pedestrian streams. Test persons with black shirts are moving from left to right (red shirts from right to left). Lanes unstable in time and space are formed. For a movie of this experiment we refer to supporting information S1.

The same phenomenon can be observed in inanimate systems, for example driven binary complex plasma or colloidal suspensions of oppositely charged particles [42,43,44]. For such phenomena, physicists use the term self-organization to express the idea that no explicit force or interaction between particles leads to the formation of lanes. Instead, volume exclusion and collisions due to the driven motion in different directions, combined with collective pushing of particles moving in the same direction, induce a segregation into lanes [42]. The occurrence of the same phenomenon in inanimate and human systems indicates that lane formation itself does not need social or psychological explanations. In this sense, lane formation is a collective phenomenon. However, it should be noted that at some places, such as very populated facilities for public transport, lane formation is organized explicitly via direction signs ("keep left"). The resulting lane formations should be conceptualized differentially from the phenomenon described above.

The second class – collective behavior – includes those phenomena which need to be understood as human or animal behavior in addition to or as a replacement for dynamics representable in physical terms. Human behavior is best described – in the tradition of behaviorism – in terms of a stimulus-response model. Stimulus-response connections can either be innate (instinct, reflex) or learned and thereby reach a very different level of complexity. To understand the relationship between a stimulus and a response one might – in the tradition of cognitive psychology – have to look at processes of perception and cognition. This category of behavior includes those areas of information processing one is not aware of. The concept of behavior is especially useful for capturing those changes in movement that humans or animals make to adapt to a certain environmental change, such as speeding up in order to stay in a group of people or steering to avoid collisions with others. Swarms and flocks serve as examples of collective behavior [45,46,47,48,49]. Mechanisms ascribed above to collective phenomena – such as volume exclusion, collision probabilities and segregation into lanes – do not suffice to describe the synchronized and flexible movements of animals in swarms or flocks. Instead, basic individual models for swarms and flocks introduced in [34,35,47,48,49] need a stimulus-response mechanism representing individuals who perceive other individuals and align their movements.

Above, we have categorized lane formation as a collective phenomenon, and even minimal models for pedestrian dynamics, such as force models with repulsive interactions [1,2,3,33,55], are able to reproduce lane formation. However, unlike inanimate particles, pedestrians form lanes to prevent collisions by steering mechanisms. Steering is a stimulus-response mechanism typical of living beings. Lane formation can therefore be viewed from two perspectives. In general, it is a collective phenomenon. If, however, the type of interaction leading to the formation of lanes is a stimulus-response mechanism, lane formation can be conceptualized as collective behavior.

Collective action as the third and most complex class must account for subjective meaning in order to understand why and how people act. Actions are chosen from a number of alternative options and are defined as those activities that actors execute intentionally and themselves regard as meaningful [61]. Even actions which are performed unconsciously at first can be rendered conscious and made understandable. Sociological and psychological action theories distinguish different kinds of actions. Two categories are very prominent [40]. Goal-oriented actions are strategic activities performed in order to reach a certain goal on the basis of situational knowledge (i.e. opening the window in order to get fresh air). Norm-oriented actions are performed in order to fit into a community and its norms (i.e. tipping waiters). Strictly speaking, actions can only be identified via self-descriptions. That someone has opened the window in order to get fresh air can only be taken for granted if the actor confirms this intention. One can also open the window in order to hear a car arriving etc. In practical terms, actions are often identified via observation and interpretation by someone who is culturally and socially able to adopt the actor's perspective.

The following example needs to be conceptualized as collective action and collective phenomenon. In emergency evacuations, clogging can occur in front of congested exits. For examples of reports of such disasters in life-threatening situations in assembly buildings we refer to [50,51]. Necessary conditions for such blockages are a limited resource leading to congestion in front of the exit in combination with a high motivation of the people to pass through the exit. For experiments with humans subjects on the influence of motivation for clogging see [52,53,54,55,56,57,58,59,60].

Again, the very same phenomena can occur in systems with inanimate particles such as granular media flowing through hoppers and silos. Clogging results from arches triggered by different forces like external forces driving particles to move in one direction, friction or contact forces between the particles as well as between particles and walls [57]. The frequency and strength of the blockages depend on the geometry of the orifice, the size and shape of the particles, the strength of the external force and many other factors [57]. Zurigel et al. compare external forces for granular media, sheep (by experiments) and human systems (by modeling), and argue that all three are equivalent [59]. That clogging occurs in inanimate and human systems leads to the conclusion that the formation of arches and the resulting clogging are collective phenomena.

While in inanimate systems the force driving the particles is external, the force in pedestrian systems is internal and is related to cognition, emotion and motivation. Even if physical models are able to represent different degrees of motivation they do not capture how the situation is perceived and how the perception changes the motivation and thus the strength of the force driving the particles to the exit. Therefore, a combination of two types of models is necessary to predict clogging: a physical model for the particles forming arches and leading to the clogging as well as a model for the motivation of different human subjects in various situations.

Clogging and lane formation illustrate the fact that collective phenomena, collective behavior and collective action do not designate strictly distinct events. Sometimes the same event can be looked at from different perspectives. Looking at goals, motivations or norms usually implies an action-oriented framework. The resulting movement might be executed as an automatic stimulus response and would therefore fall into the category of behavior. Movements might even result in phenomena for whose description no psychological variables are needed at all. Therefore, the adequate choice of one of the three concepts partly depends on the research question. Following the principle of Occam's razor, the least complex framework should be applied first and extended if necessary.

# Empirical data part I: Pedestrian dynamics in two entrance setups. An experimental study

Large gatherings are challenging for organizers and authorities. The research project BaSiGo developed safety and security modules for such events. As a part of this project, the origins of critical conditions due to overcrowding were studied in June 2013. Over four days, experiments with more than 2000 test subjects were performed in a hall at the exhibition site in Düsseldorf, Germany. Participants were acquired by advertisements at nearby universities. Thus the test subjects were mainly adults between 20 and 30 years old. One of these experiments focused on entrances to music events where pushing by highly motivated fans can lead to dangerous situations.

## Methods

The experiment shown in Figure 2 or movies S2 and S3 studied the influence of the spatial arrangement of the barriers on the behavior of participants. An experienced crowd manager (who plans and coordinates entrance and exit situations of large events) suggested different setups for the experiment, two are examined here. In the first setup, no guiding barriers were installed and participants were positioned loosely in a half circle in front of the two entrances. For the second setup, barriers were arranged to form a corridor, guiding the test persons in a limited space from the side to the two entrances. The participants were positioned loosely inside the corridor. In both setups, the test subjects were advised to imagine an entry situation to a concert of their favorite artist. They were told that the crowd was about to be admitted and that they should try to be one of the first to pass through the entrance. Upon a command by the investigator, the entrances were opened and the participants started to enter. All participants signed a written informed consent and agreed to publication of pictures and videos shown in this article. Only anonymous data were used for the experiments. The Federal Ministry of Education and Research (Germany) has approved the project. No ethical concerns were mentioned.

# Results

A visual inspection of the snapshots already shows that different spatial structures of the barriers lead to significantly different densities in front of the entrances. In the following section the experiment is analyzed in detail.

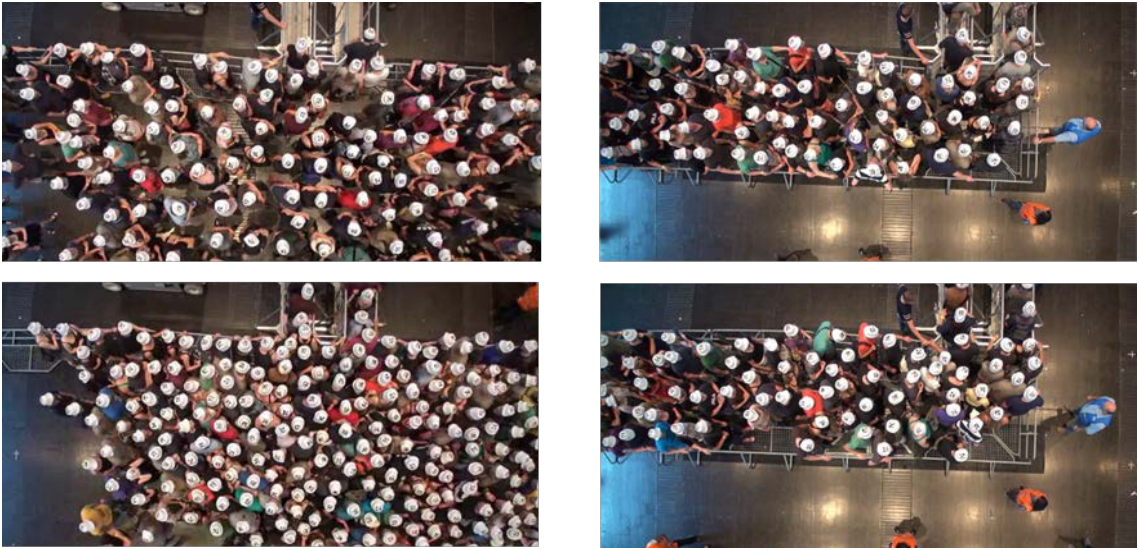

**Fig 2.** Left: Entry without guiding barriers (semicircle setup). Test subjects are positioned in a semicircle in front of two entrances 0.5 m in width. Right: Entry with guiding barriers (corridor setup). From top to bottom t = 0 and 6 s after the command to start entering. The densities in front of the barriers at t=0 are comparable. With guiding barriers, the density in front of the entrances is significantly smaller at t = 6 s. For movies of the experiments we refer to supporting information S2 (semicircle setup) and S3 (corridor setup).

The footage was processed with PeTrack [62], a software for automatic extraction of pedestrian trajectories from video recordings. PeTrack offers methods for calibrating the data and relating pixel coordinates with world coordinates. The extraction of the trajectories is achieved by recognizing and tracking the white caps as markers. For image processing, it is defined as a coherent area of white pixels with a specific surface and extension. The exact position is presented as the center of the ellipse that approximates the border of the white marker. The QR code is neglected. For more details we refer to [62]. The resulting trajectories were corrected manually with an emphasis on removing false positives (which appeared, for example, because of pedestrians wearing white clothes).

Processing with PeTrack results in trajectories for each detected person defined as a sequence of positions $p_{id}(t_i) = (x(t_i), y(t_i))$ at time $t_i$ with $id$ as a unique number identifying the person. The frame rate giving the time interval and the frequency of the sequence was 25 Hz. The trajectories represent positions of the head projected on the ground.

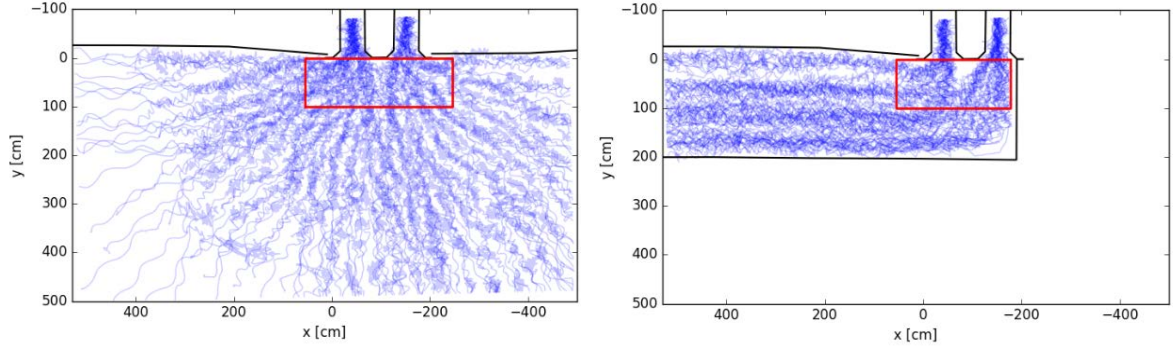

**Fig 3.** Cumulative trajectories for the semicircle (left) and corridor (right) setups. The red rectangles show the measurement areas for the time series of the density in front of the entrances. Trajectories in the corridor setup show four lanes in the straight part and merging at the entrances.

In Figure 2 the extracted trajectories for both setups are shown. The sequence $p_{id}$ is plotted for each pedestrian. Due to the bipedal locomotor system and the tracking of the head, the trajectories oscillate along the center of mass movement of the body. This effect becomes stronger with decreasing speed and is maximal if the person is waiting on the same spot moving their center of mass from one leg to the other. The trajectories are plotted with a low opacity to let spatio-temporal structures emerge. For the semicircle setup, a star-like structure appears indicating the symmetry of movements to the center where the entrances are located. For the corridor setup, four lanes appear indicating the formation of an ordering structure.

To quantify the differences in density we use the Voronoi method introduced in [63]. A Voronoi cell includes all points in space that are nearer to the person $id$ then to any other and its area $A_{id}$ is determined for each person. The Voronoi cell allows the density distribution of that space to be calculated as

$$\rho_{xy} = \frac{1}{A_{id}} \quad \text{if } (x,y) \in A_{id} \tag{1}$$

The density for an arbitrary measurement area containing several Voronoi cells is obtained by the equation above where $A_{MA}$ is the size of the measurement area.

$$\langle \rho \rangle_v = \frac{\iint \rho_{xy} dxdy}{A_{MA}} \tag{2}$$

A rectangular measurement area of 3 by 1 meters, centered right in front of the two entrances, was chosen for the semicircle setup. For the corridor setup, the area was slightly truncated to 2.3 by 1 meter to fit into the boundaries of the corridor. The time dependence of the density for the first 120 seconds is shown in Figure 3.

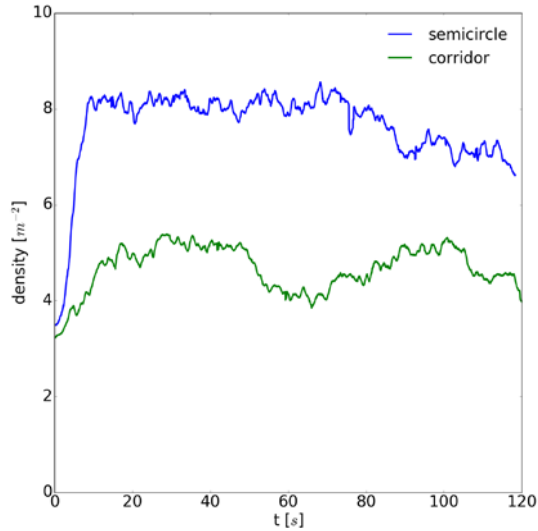

**Fig 4.** Time series for density in front of the entrances. After the start command, the constriction effect leads to an increase of density in front of the entrances. For the semicircle setup the density fluctuates around 8 m$^{-2}$ after the constriction, while for the corridor setup it is around 5 m$^{-2}$.

For the semicircle setup, the density increases sharply starting from 3.8 and reaches 8 m$^{-2}$ within the first 10 seconds. After the start signal, all the people move as close as possible to the entrances until the free spaces are filled. This so-called constriction effect brings movement nearly to a standstill. For the corridor setup, this effect is distinctly smaller leading to a density fluctuating around 5 m$^{-2}$ after 20 seconds.

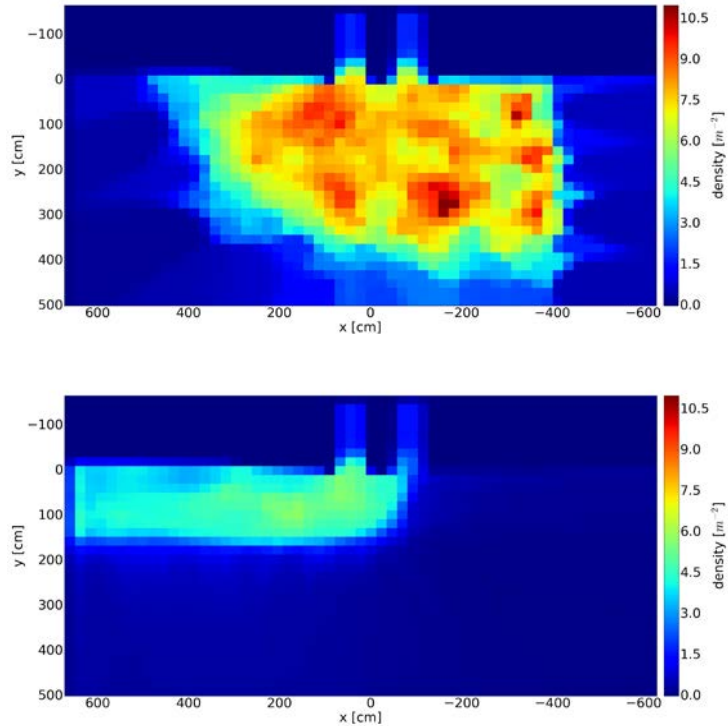

**Fig 5.** Density profiles as mean values for time interval t=20s to 80s, see Figure 4. In the semicircle setup (top) a heterogeneous structure with hot spots up to 11 m$^{-2}$ appears.

Density profiles are calculated for a spatial analysis, see Figure 4. The area is divided into tiles with a size of 0.2 by 0.2 m and the densities in the tiles are averaged over the steady state ranging from 20s to 80s. For each tile and for each frame within the steady state, the Voronoi density was calculated, added up (per tile) and averaged over time. The local density for the semicircle setup reaches values of 11 m$^{-2}$, while it does not exceed 5.7 m$^{-2}$ for the corridor setup. The highest density for the semicircle setup does not occur right in front of the entrance. Instead, several hot spots appear, distributed in the center. Some of the hot spots can potentially be explained by groups of people standing very close together or individuals pushing vigorously. In the corridor setup not only is the maximal density significantly lower but also the spatial distribution is free of hot spots and is homogenous in contrast to the semicircle setup.

Finally, we study the relation between time to entrance and distance to entrance. For this analysis, we only use the trajectory data of those persons who reached the entrance during the runtime of the experiment. For each person, the Euclidean distance between the current position and the lower center point of the entrance chosen (left or right) is determined. The time to entrance is given by the time the person needs to arrive at the entrance. This calculation was done for the initial position at t=0s (highlighted as large dots in light blue) and every 10th frame of each trajectory, see Figure 4a. Due to differences between the initial conditions in the two experiments, we neglect the first 6 seconds of the experiment in the analysis, so that the entrances are filled with people, see Figure 1.

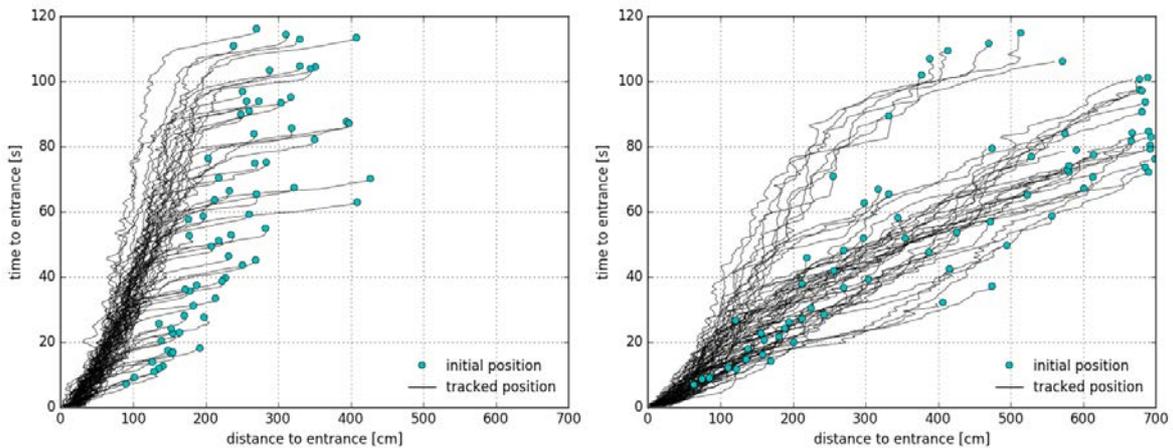

**Fig 6.** Relation between time and distance to entrance for the semicircle setup (left) and corridor setup (right). Light blue dots correspond to the initial position while black lines show the cumulative data for the relation during entering.

For the semicircle setup, again two phases become obvious. First, the constriction phase indicated by a strong reduction of distance to the entrance and a small reduction of time to entrance. After the constriction, the phase distance decreases more slowly. The corridor setup does not show any constriction and displays a more homogeneous trend for the relation. It is noteworthy that two bunches appear in the corridor setup merging at distance of 1.5 m and 40 seconds. People in to the upper bunch need much longer to enter. An examination of the film shows that people in the lane located at the inner boundaries are blocked by the stream entering the left entrance thus leading to longer waiting times for that lane. The reason is that people entering from the front have an advantage due to the straight entrance and do not leave any space for people from the side to filter into the stream.

To examine the relation between distance and time to entrance in detail, a linear regression is performed on the initial values and the data of the time interval from 20 seconds to 80 seconds, see Figure 6. This partition was chosen to analyze the constriction phase separately from the phase when the high density in the semicircle setup had already been established. Speed during the entering progress can be estimated by the slope of the regression line. A strong linear relation indicates that people standing nearer to the entrance enter earlier. In this sense, the correlation coefficient is a measure of the justness of the process.

**Table 1.** Results of a linear regression of the relation between time and distance to entrances. While the slope as an indicator of the speed of the entering progress is different the correlation coefficient as an indicator of the justness of the process shows negligible variations.

|  | Semicircle | | Corridor | |
|---|---|---|---|---|
|  | Slope | Corrcoef | Slope | Corrcoef |
| Initial positions (t=0s) | 0.32 [s/cm] | 0.79 | 0.12 [s/cm] | 0.81 |
| Time interval [20 s,80 s] | 0.51 [s/cm] | 0.91 | 0.13 [s/cm] | 0.82 |

Due to the constriction, the slope of the semicircle setup increases for the time interval between 20 and 80 seconds in comparison to the initial positions. After the constriction ends, the slope is 0.51 s/cm thus giving a progress speed of around 2 cm/s for the semicircle setup. For the corridor setup, the progress speed does not change in the course of the experiment and is remarkably higher amounting to 8.3 cm/s. Surprisingly, the values of the correlation coefficients show no strong differences. Both setups seem to be equally just, with the value of the correlation coefficients near to one. One reason for this unexpected uniformity could be the unjustness in the corridor setup due to the lane which is discriminated against at the inner boundary. A further explanation for the high correlation in the semicircle setup is that options for cheating by pushing or overtaking are rare after the constriction has taken place. Instead, in the corridor setup where density is lower some people give themselves an edge by overtaking.

To sum up, the qualitative differences in the setups conform to observations in everyday life. Density is much higher in the semicircle setup (11 m$^{-2}$) than in the corridor setup (5.7 m$^{-2}$). The high density in the semicircle setup is a result of a constriction effect. After the start signal, test subjects move forward filling up the space quickly so that no gaps exist. By linear regression, we found that the progress speed is much slower in the semicircle setup (2 cm/s) than in the corridor setup (8.3 cm/s). The justness measured by the strength of the linear correlation is – in contrast to our expectation – not different. In both setups, people who are closer to the entrance exit earlier.

# Empirical data part II: Perception and evaluation of entrance scenarios. A questionnaire study

Participants in this study watched pictures and videos of the two entrance setups described above (semicircle and corridor). The questionnaire addresses five aspects of their perception and evaluation: 1. justness, 2. level of comfort, 3. likelihood of being one of the first 100 to access the concert, 4. social norms.

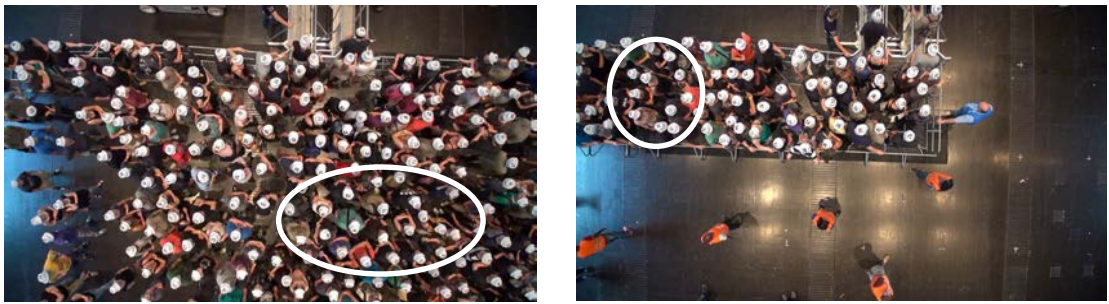

**Fig 7.** Semicircle (left) and corridor (right) setups. Participants are instructed to imagine that they are standing in the designated areas (white ellipses).

## Methods

### Participants

N = 60 participants were recruited from undergraduate social sciences (n = 48) or engineering (n = 12) classes in Germany. None of them had taken part in the experimental study described above (empirical data part I). The participants' mean age was 25 (SD = 4.1). All participants signed a written informed consent. Identification data from the informed consent were kept separate from the questionnaire throughout the entire procedure.

### Procedure

Pictures and videos of entrance scenarios were projected onto a screen and played without sound. Participants were instructed to imagine they were part of the group and standing in one of the designated areas (Fig 7). They were asked to envision an entrance to a concert by their favorite band at which only the first 100 people were able to access the concert.

The questionnaire was applied at two points of time: when participants were looking at a freeze frame of the moment just before the entrance procedure started (termed "before video" below) and after they had observed people accessing for two minutes (termed "after video" below).

This questionnaire study uses a within-subject design. Therefore, all the participants watched videos of both entrance setups (semicircle and corridor). In order to control the sequence effects, half of the participants watched the videos in the reverse order. In total, the questionnaire was applied four times: Before and after watching the semicircle setup, before and after watching the corridor setup.

## Questionnaire

The questionnaire (originally in German) contains four main items: 1) How just is this entrance procedure? (6-point scale, 1=very unjust, 6=very just), 2) How likely is it that you will be one of the first 100 who are able to access the concert? (6-point scale, 1=very unlikely, 6=very likely), 3) How comfortable do you feel? (6-point scale, 1=very uncomfortable, 6=very comfortable), 4) Can you contribute to accessing the concert faster? (yes/no). In addition to question 4, strategies for being faster were requested (open-ended question).

The questionnaires administered after watching the videos contain two additional questions: Are you under the impression that you are making progress towards the entrance? (6-point scale, 1= not at all, 2= very much). Have you observed anyone behaving unfairly or inappropriately? (1=nobody, 2=few people, 3=many people, 4=everyone) And what kind of inappropriate behavior have you observed? (open-ended question). At the end, participants are asked to name social norms for the semicircle and corridor setups (open-ended question). The questionnaires are completely anonymous and cannot be traced back to the participants. For the questionnaire study no ethical approval was obtained because the questions are not related to any personal or intimate information but rather ask for an evaluation of crowded, non-violent scenes on video which are regularly observable in everyday life.

## Statistical analysis

Two-way repeated measures analysis of variance (setup × point of time) was conducted to evaluate perceived justness, likelihood of access, comfort and progress. Items on nominal level (possibility of contributing to faster access, observed inappropriate behavior) are provided in percentages. Significance was set at $p < 0.05$. All statistical analyses were performed using SPSS for mac, version 23. Answers on open-ended questions were slightly generalized with regard to content, translated and counted. For the purpose of this article only the three most frequently used answers are presented.

## Results

*Justness:* Participants in the questionnaire study perceive the corridor setup as being significantly juster than the semicircle setup, irrespective of the time of evaluation ($F_{1, 58} = 150.7$ ; $p < 0.0001$). The time of evaluation (before and after watching) and setup (semicircle and corridor) interact significantly ($F_{1, 58} = 6.84$; $p = 0.011$): Whereas in the corridor setup participants perceive the situation as being even juster after observing the experiment, in the semicircle setup it is perceived as less just. As a result of this interaction, the total perceived difference of justness increases after the video has been shown, see Figure 8 (left). The time of evaluation by itself does not change the evaluation of justness: Overall, observing the experiment does not significantly change the evaluation of justness ($F_{1, 58} = 0.58$; $p = 0.451$).

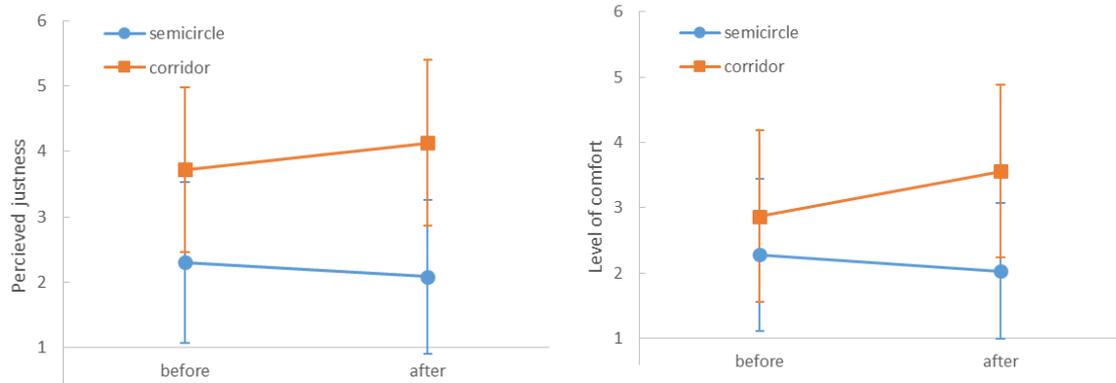

**Fig 8.** Perceived justness: mean, standard deviation (left). Corridor setup is perceived as significantly juster than the semicircle setup. After watching the video, the difference in perception is confirmed and increases. Level of comfort: mean, standard deviation (right). Participants feel more comfortable in the corridor setup than in the semicircle setup. Again, the difference is confirmed after watching the videos and increases.

*Likelihood of being one of the first 100 in the audience:* In the corridor setup, participants find it more likely that they would be one of the first 100 people able to access the concert ($F_{1, 58}$ = 42.43; $p < 0.0001$). Time of evaluation by itself does not change the perceived likelihood of success ($F_{1, 58}$ = 0.13; $p = 0.724$). Time of evaluation and setup do not interact significantly ($F_{1, 58}$ = 3.65; $p = 0.061$).

*Level of comfort:* Participants feel more comfortable in the corridor setup ($F_{1, 56}$ = 57.52; $p < 0.0001$). Furthermore, the time of evaluation shows a significant main effect ($F_{1, 56}$ = 5.96; $p = 0.018$). As can be seen in Figure 8 (right), this main effect can be explained by the strong increase of the level of comfort in in the corridor setup. Finally, time of evaluation and setup interact significantly ($F_{1, 56}$ = 13.51; $p = 0.001$), meaning that the difference between the semicircle and corridor setups increases after observing the entrance procedure.

*Perceived progress*: After watching the videos, participants were asked whether they perceive any progress towards the entrance. In the corridor setup, participants perceive more progress (mean 2.56 ± 0.99) than in the semicircle setup (mean 3.78 ±1.09) with ($F_{1, 58}$ = 52.17; $p < 0.0001$).

None of these results was influenced significantly by the order in which participants watched the videos (semicircle or corridor setup as the first video).

*Inappropriate behavior:* In the semicircle setup, participants observe far more people showing inappropriate or unfair behavior, see Table 2.

**Table 2. People observed showing inappropriate behavior**

| People showing inappropriate behavior | Semicircle | Corridor |
|---|---|---|
| Nobody/few | 59.4 %. | 94.9 % |
| A lot/all | 40.6 % | 5.1 % |

Participants list the following forms of inappropriate behavior:

**Table 3. Forms of inappropriate behavior (frequency of occurrence, three most frequently mentioned only)**

| Semicircle | Corridor |
|---|---|
| - pushing and shoving (35)<br>- pushing someone aside (11)<br>- jostling (9)<br>- … | - pushing and shoving (16)<br>- slightly pushing and shoving (4)<br>- jostling (3)<br>- standing still (3)<br>- …. |

*Individual contribution to faster access*: In the semicircle setup, slightly more participants think that they can contribute to faster access. This holds true for both points in time: before and after watching the videos, see Table 4. For both setups, the number of participants who think that they can contribute to faster access decreases strongly after watching the videos. Table 5 lists the strategies for faster access for each setup and point in time.

**Table 4. Percentage of participants who think that they can contribute to faster access**

|  | Semicircle | Corridor |
|---|---|---|
| Before | 78.3 % | 71.7 % |
| After | 48.3 % | 31.7 % |

**Table 5. Strategies for faster access (frequency of occurrence, three most frequently mentioned only)**

|  | Semicircle | Corridor |
|---|---|---|
| Before | - Pushing and shoving (25)<br>- Using and filling gaps (10)<br>- Using elbows/arms/shoulders (9) | - Pushing and shoving (21)<br>- Staying on the left hand side (11)<br>- Using and filling gaps (4) |
| After | - Pushing and shoving (14)<br>- Using central area, avoiding flanks (3)<br>- Using elbows/arms/shoulders (3) | - Pushing and shoving (7)<br>- Turning to the right hand side (6)<br>- Staying on the left hand side (5) |

*Social norms*: Participants apply the following social norms to both setups:

**Table 6. Social norms (frequency of occurrence; three most frequently mentioned only)**

| Semicircle | Corridor |
|---|---|
| • The strongest wins/right of the stronger (15)<br>• No rules (15)<br>• First come, first served (7) | • Norm of queuing/lining up (16)<br>• Orderly behavior (11)<br>• Pushing and shoving are forbidden (10) |

# Discussion

Participants clearly evaluate the corridor setup more positively than the semicircle setup: it is seen as more comfortable, juster and progressing faster. Furthermore, participants find they are more likely to be successful in the corridor setup. Less inappropriate behavior is observed here. In both setups pushing, shoving and jostling are evaluated as inappropriate.

Normatively, the semicircle setup is described as a situation without clear social norms in which the principles "first come first served" and "right of the stronger" apply. On the other hand, the corridor setup is perceived as an orderly situation for which the norm of queuing applies.

In the semicircle setup more participants think that they can contribute to faster access than in the corridor setup. However, in both setups and at both points in time they mention pushing and shoving as the number one strategy for faster entry. It is interesting to note that pushing and shoving are at the same time seen as inappropriate while being listed as promising strategies.

# Conclusion and outlook

Combining the results from the experiment and questionnaire, the study reveals two important analogies but also two interesting discrepancies: 1. As expected, the higher density in the semicircle setup is accompanied by a much lower level of comfort. 2. Furthermore, perceived progress speed towards the entrance and real walking speed are higher in the corridor setup. 3. Participants evaluate the corridor setup as juster, but measured objectively, both setups are equally just with the tendency of the semicircle setup to be juster. This can be interpreted as follows: In the corridor setup, most participants rely on the norm of queuing and expect to enter when it is their turn. As a result they keep a distance from others and leave gaps unused. This allows a few participants to cheat and overtake. In addition, the left lane is disadvantaged. In the semicircle setup, participants expect the procedure to be unjust and competitive. Therefore, they all move forward, close gaps and increase density. This makes overtaking almost impossible and results – surprisingly – in a just entrance procedure. 4. In the semicircle setup, more participants think that they can contribute to accessing faster, even after they have watched the video. This assessment is not in accordance with the high densities in the semicircle setup, which leaves practically no space for individual contributions. In interpreting the results, it should be kept in mind that participants in the questionnaire study did not actually take part in the entrance experiment. Further research is needed to capture the perceptions and evaluations of people who are part of the crowd at a specific time.

The main effect to be explained is the constriction effect in the semicircle setup and its absence in the corridor setup. The constriction effect is predictable with physical models. A similar dynamic is to be expected if humans are replaced by animals or even inanimate particles. Following the systematization developed above, the constriction effect can therefore be categorized as a collective phenomenon. However, the absence of the constriction effect in the corridor setup would only be expected in systems with humans. It can be explained by the social norm of queuing and a more passive strategy for entering the concert hall (including waiting one's turn and not expecting to be able to influence speed). The dynamic observed in the corridor setup should therefore be classified as collective action, participants act intentionally and in a norm-oriented manner.

All individual-based models from the natural sciences use intended velocity as an external parameter. In using such models, the two setups in our study would have to be treated separately by manually adjusting the intended velocity. However, in order to predict the differences between the two setups and the occurrence of queuing a model is needed which starts one step earlier and treats intended velocity as a quantity which has to be modeled itself, as in [6,64]. From the results presented here, this additional step in modeling must go beyond merely considering physical characteristics and needs to integrate subjective perception and associated strategies and social norms. Following Lewin's field theory [65], the subjective perspective mediates the influence of physical characteristics on action. Referring to the theoretical framework of this article, existing models from the natural sciences are appropriate for modeling collective phenomena and behavior but need to be expanded by including sociological and psychological concepts to model collective action. Vice versa, physical concepts such as density, flow and speed are helpful in describing whether options exist for behaving or acting as a human subject and where actions can no longer be executed because they are limited by physical conditions. Referring to our data in the semicircle setup, people's options to act are strongly limited after the constriction has taken place because it is too dense. Thus, natural and social psychology truly complement each other in their perspective on crowd dynamics.

## Acknowledgments

We thank Maik Boltes for his valuable support in extracting the trajectories. The empirical data analyzed in the first section were part of the BaSiGo project. Armin Seyfried would like to thank the team responsible for the planning and execution of the experiment, in particular Martin Hoube. The project "Basigo – Bausteine für die Sicherheit von Großveranstaltungen" was funded by the German Federal Ministry of Education and Research under grant number: 13N12045.